\documentclass[mathleft
]{an}
\usepackage{graphicx}
\usepackage{times}
\usepackage{natbib} 
\usepackage[hyperindex,breaklinks]{hyperref}

\begin{document}

\Pagespan{1}{}
\Yearpublication{2009}%
\Yearsubmission{2005}%
\Month{11}%
\Volume{999}%
\Issue{88}%


\title{Sigma-drop in galaxies and the sigma-metallicity degeneracy}

\author{Mina Koleva\inst{1,2}\fnmsep\thanks{Corresponding author:
  {mina.koleva@obs.univ-lyon1.fr}}
\and  Philippe Prugniel\inst{1}
\and  Sven De~Rijcke\inst{3}
}
\titlerunning{Sigma-drop in galaxies}
\authorrunning{M. Koleva, Ph. Prugniel \& S. De Rijcke}
\institute{Universit\'e de Lyon, Lyon, F-69000, France ; Universit\'e Lyon~1,
Villeurbanne, F-69622, France; Centre de Recherche Astronomique de
Lyon, 
Observatoire de Lyon, St. Genis Laval, F-69561, France ; CNRS, UMR 5574 ;
\and 
Department of Astronomy, St. Kliment Ohridski University of Sofia, 5 James
Bourchier Blvd., BG-1164 Sofia, Bulgaria
\and 
Astronomical Observatory, Ghent University, Krijgslaan 281, S9, B-9000 Gent, Belgium
}

\received{15 Sept 2008}
\accepted{15 Sept 2008}
\publonline{later}

\keywords{Stellar populations -- Internal kinematics -- Metallicity -- Galaxies }

\abstract{ In some galaxies, the central velocity dispersion,
  $\sigma$, is depressed with respect to the surroundings. This
  sigma-drop phenomenon may have different physical origins, bearing
  information about the internal dynamics of the host galaxy. In this
  article, we stress the importance also of observational artifacts
  due to the $\sigma$-metallicity degeneracy: when a spectrum of a
  population is compared with a template of miss-matched metallicity,
  the velocity dispersion may be wrongly estimated. A sigma-drop may
  appear in place of a metallicity peak.  The discussion is
  illustrated using VLT/FORS spectra of diffuse elliptical
  galaxies. Some of the sigma-drop galaxies reported in the literature
  may be analysis artifacts.  }

\maketitle

\section{Introduction}

In a fraction of galaxies, mostly spirals but also ellipticals, the
velocity dispersion is depressed in the center \citep[e. g.][]{comeron08}. 
However, simple density profile models, like the de Vaucouleurs profile,
predict a continued increase inwards.
In the very center, approaching the central massive black hole, a further
steep increase is even expected \citep{DR90}.

\begin{figure}
\includegraphics[width=8cm]{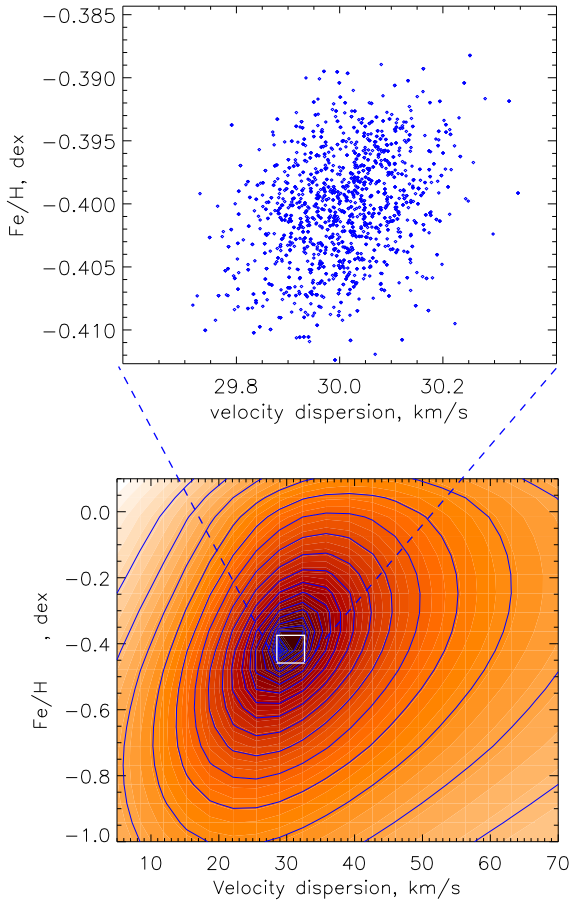}
\caption{
Monte Carlo simulations and $\chi^2$ map. The bottom panel shows a $\chi^2$ map
projecting the parameter space on the $\sigma$ vs. metallicity plane. For
each point of a grid, the other free parameters, cz, age and
multiplicative polynomial are optimize and the map shows the variations
of the resulting  $\chi^2$ values. The iso-$\chi^2$ contours are elongated 
along a direction mixing the two parameters. The top panel shows a Monte-Carlo
experiment, simulating the effect of a Gaussian noise comparable to the
FORS observations (S/N=70). Each of the 3000 points represent the analysis
of a different realization of the noise. The cloud is elongated in a 
diagonal direction, showing the $\sigma$ - metallicity degeneracy.
}
\label{fig:1}
\end{figure}

These observed $\sigma$-drops may have various physical
explanations. They may indicate the absence of a central mass
concentration \citep{DR90} or more generally a flat
density core. The latter is certainly expected in diffuse elliptical
galaxies (dE) with shallow density profiles, flatter than
exponential. Such examples may be found in \citet{SP02}.
Still, in the list of $\sigma$-drop galaxies, the spiral galaxies
are the majority and since the stellar kinematics is less studied in spirals
it is quite probable that most of the $\sigma$-drops occur in spirals.
Some studies suggest up to 50 \% of $\sigma$-drops among the spirals
\citep{comeron08}.

In spiral galaxies, a $\sigma$-drop may result from a young stellar
population formed from a rotating gas disk funneled toward the center
by the dynamical action of a bar \citep{wosniak03}.  Simulations
show that this feature may persist 1 Gyr until the newborn population
mixes with the underlying population and its luminosity fades. A
sustained gas supply and star formation may produce a long-lived cold
stellar disk.

In this article we investigate another phenomenon affecting the detection
of $\sigma$-drop galaxies: the $\sigma$-metallicity degeneracy.

\section{Analysis of line-of-sight integrated spectra}

To study the internal kinematics of a stellar population, an
observation is compared to a template (an observed star or a
population model) having the same intrinsic broadening (due to the
spectrograph) and no physical broadening.  The relative broadening
between the two is the physical velocity dispersion.  Various
algorithms for measuring the internal kinematics, or line-of-sight
velocity distribution (LOSVD) are available \citep[for example][]{CE04}. 
They search for the optimal convolution kernel
(Gaussian or Gauss-Hermite expansion) that allows to transform the
cold template into the observation. But the historical algorithms,
cross-correlation \citep{TD79}, Fourier quotient \citep{sargent77}
or their variants as the Fourier correlation quotient
\citep{bender90} have difficulties to handle properly the noise and
outliers in the signal. Preferred methods, nowadays, perform the fits
in pixel space \citep[e. g.][]{CE04}.

Whatever the algorithm that is used, one of the key issues is to find
a template that matches the observation. It is known for a long time
that the determined velocity dispersion depends of the spectral type
and metallicity of the stellar template \citep[e. g.][]{bender90}.  Some
studies use different stellar templates and average the results, this
also washes out variations of the intrinsic resolution of the
spectrograph. Others try to properly match the template, in particular
its metallicity, or to use a linear combination of several templates
(optimal template fitting).

The templates are commonly stars observed using the same setup as for
the galaxy.  But is is also possible to use spectra from stellar
libraries, or even synthetic population templates \citep{vazdekis99}. This
suppresses the need to observe a large number of reference stars, but
adds the difficulty of injecting in the template spectrum the 
line-spread function (LSF) of the spectrograph relative to the template
\citep{kol_pru08}.

In the present article we are using the ULySS package \citep[soon available
on-line][]{ulyss} that fits a spectrum in pixel
space. This package contains the tools to analyze the relative LSF,
and it matches the observation with either a stellar or
stellar-population spectrum after injecting this LSF. In this work, the
fit will be made using population models -- PegaseHR, \citep{PHR} 
with Elodie.3.1 \citep{PS01, PS07} -- 
parametrized with their age and metallicity. Hence, the
LOSVD ($cz$ and $\sigma$), age and metallicity are simultaneously
determined.

\begin{figure}
\includegraphics[width=8cm]{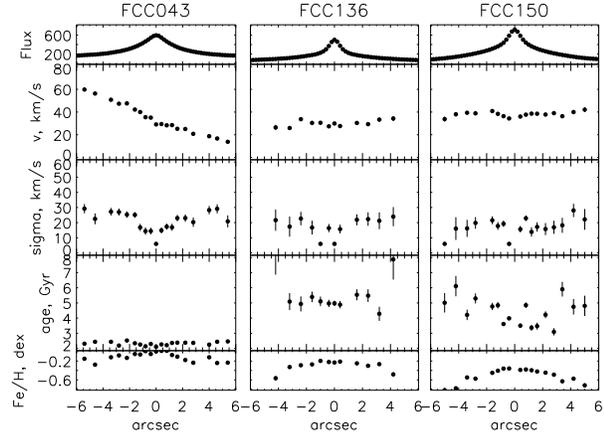}
\caption{Radial profiles of three galaxies, from left to right 
-- FCC043, FCC136 and FCC150.
From top to bottom are shown flux, radial 
velocity, $\sigma$, SSP-equivalent age and metallicity as a function 
of the radius. 
}
\label{fig:2}
\end{figure}

This optimization is made for each spectrum. It means that for a long-slit 
observation we measure at once the kinematical and population profiles 
(Figure~\ref{fig:2}).

\section{The $\sigma$-metallicity degeneracy}
It is known, or at least suspected, for a long time that the
metallicity mismatch biases the $\sigma$ measurement \citep{laird85}, 
but this effect has been widely overlooked. The
$\sigma$-metallicity degeneracy was studied in \citet{kol_bav07}
and the magnitude of the effect was established from Monte-Carlo
simulations. The intuitive explanation for this effect is that if the
metallicity of the template is lower than that of the galaxy, its
metallic lines are less deep than those of the observations, and the
minimization program compensates the mismatch by decreasing $\sigma$
(which in turn increases the depth of the absorption lines). Figure~\ref{fig:1}
presents a Monte-Carlo experiment and a $\chi^2$ map illustrating this
effect. This degeneracy may be written as:

\begin{figure*}
\includegraphics[width=17cm]{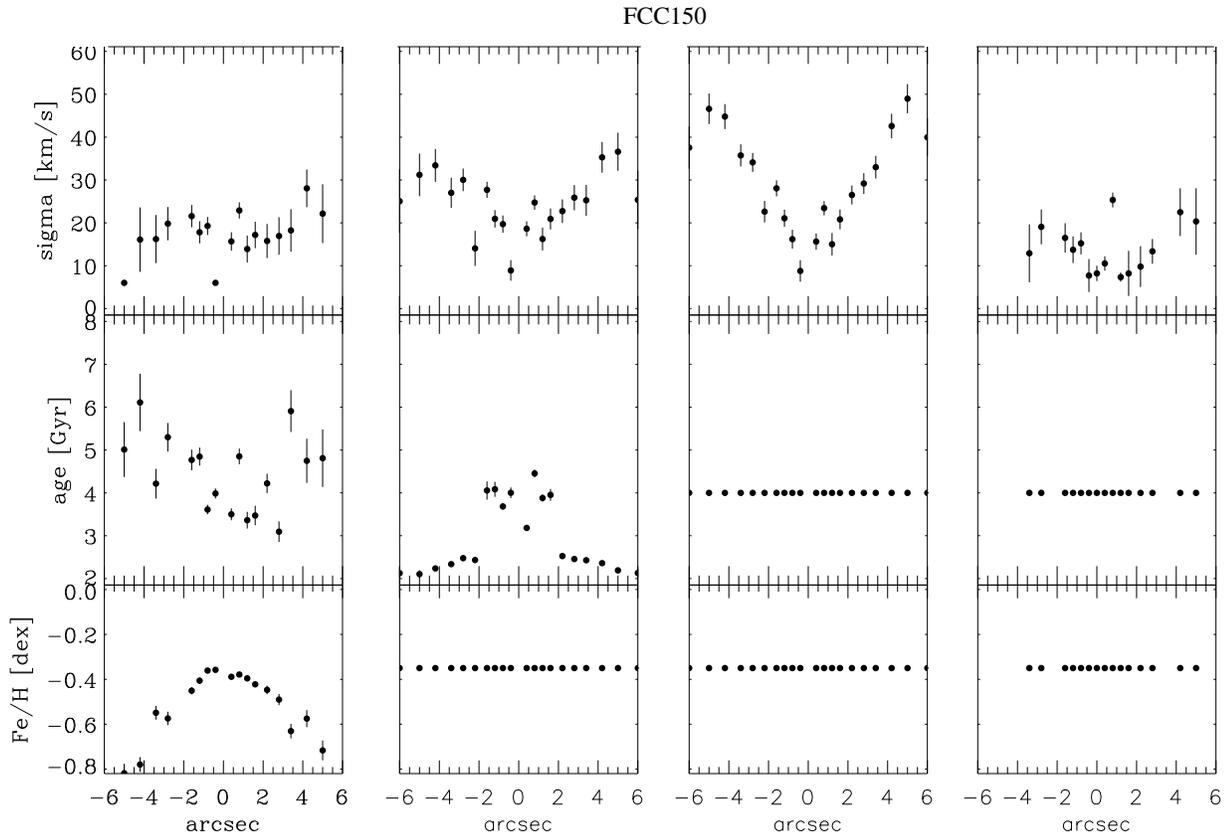}
\caption{
Radial profiles of FCC150. The four panels represent minimization with different free/fixed
parameters. From left to right -- all parameters are free; the metallicity is fixed; 
the age and metallicity are fixed, age/metallicity fixed plus an additive continuum. 
A mismatch of the metallicity (and age) could result in an artificial sigma drop.
An additive term reduce the mismatch.
}
\label{fig:3}
\end{figure*}

$$
 {\delta(\sigma)\over \sigma} \approx 0.4 \times \delta([Fe/H])
$$

Where $\sigma$ is in km/s, and $[Fe/H]$ in dex.
A 0.1 dex mismatch of the metallicity results in a 4\% error on the velocity
dispersion. This approximate formula was determined from simulations and
is valid within 30\% for a wide range of parameters. The effect is stronger
for smaller velocity dispersions and younger ages. 

If a long-slit spectrum is analyzed with a fixed template, a central
peak of metallicity will be compensated by an underestimated $\sigma$.
If the intrinsic $\sigma$ profile is shallow, the fitted profile may possibly
display a $\sigma$-drop.

To illustrate this question we analyze below high quality long-slit
observations of a sample of dEs obtained with FORS1 at the VLT.
Figure~\ref{fig:2} shows the LOSVD and population profiles in the center of 3
galaxies.  The metallicity peaks in the center, while the age is
dropping (this is an analysis with a single stellar population, the
real stellar mix is certainly more complex, see Koleva et al. in
preparation).  The $\sigma$ profile does not show prominent
systematics (note that the instrumental velocity dispersion of these
observations is 64 km/s. The reported $\sigma$ values are still
uncertain because we did not yet study the variations of the
spectrograph resolution with time and seeing).

These galaxies, with peaked metallicity, are exactly the configuration
where a single metallicity template would bias $\sigma$. To
demonstrate it, Fig. 3 repeats the analysis of FCC150 with a fixed
metallicity and then fixed age and metallicity. Fixing the metallicity
inverts the age trend in the center: it becomes a peak instead of a
drop. This is a consequence of the well known age-metallicity
degeneracy, the underestimated metallicity is balanced with an older
age. But at the same time the $\sigma$ profile displays a clear
minimum in the center. Fixing both the age and the metallicity
increases again the $\sigma$-drop. 

The last panel of Fig. 3 presents
a test where an additive polynomial was included in the model. Such a trick
is often used in the programs measuring the kinematics in order to absorb
the template mismatch. We see that this method is efficient: The
artificial  $\sigma$-drop is removed.

A classical fixed template analysis would detect an artificial 
$\sigma$-drop even with data of quite lower quality.

\section{Conclusion}

We have shown that some $\sigma$-drops may be artifacts of the
analysis method. They may result from a peak in the metallicity, which
is actually often expected in the centers of the galaxies.

This naturally does not mean that all the $\sigma$-drops are of this
nature, but that individual observations must be examined with
caution. One has to ask the questions: (i) Is the analysis made with a
fixed-metallicity template, or with an additive polynomial? 
(ii) Is the magnitude of the effect
comparable with the sigma-metal\-li\-ci\-ty degeneracy?

In the literature, we found two samples of $\sigma$-drop
galaxies:~\citet{emsellem08} and \citet{comeron08}.  Half of the 20
$\sigma$-drop galaxies were detected in various studies using
fixed-metallicity templates and the magnitude of the drops are
generally consistent with them resulting from the sigma-metallicity
degeneracy. Some of these studies are reducing the mismatch
with an additive term, but it is difficult to determine their
reliability without testing the actual analysis program.

Only recent studies, like \citet{dumas07}, \citet{emsellem01}, and
some other based on SAURON data use an optimal template fit to each
single spectrum. Unfortunately they use as a reference the \citet{jones98}
stellar library which has a limited coverage in metallicity and may
therefore not match perfectly the observations.

We conclude that the fraction of $\sigma$-drop is certainly\break
overestimated, and that the $\sigma$-metallicity degeneracy should
not be neglected.  Fitting in the same time the kinematics and the
chemical characteristics of the population is definitely the best
solution to correctly reconstruct stellar population properties.

%
\acknowledgements
Mina Koleva acknowledges the financial support of the meeting organizers.

\bibliographystyle{aa} 
\bibliography{GSD08}   


\end{document}